\documentclass{article}
\usepackage[utf8]{inputenc}
\usepackage{authblk}
\usepackage{indentfirst}
\usepackage{hyperref}
\usepackage{longtable}
\usepackage{setspace}
\usepackage[margin=1in]{geometry}
\usepackage{graphicx}
\usepackage{subcaption}
\usepackage{amsmath}
\usepackage{lineno}
\usepackage{multirow}
\usepackage{color,soul}

\usepackage[biblabel]{cite}

\title{Performance of the r$^{2}$SCAN functional in transition metal oxides}

\author[1]{S. Swathilakshmi}
\author[1]{Reshma Devi}
\author[1,*]{Gopalakrishnan Sai Gautam}

\affil[1]{Department of Materials Engineering, Indian Institute of Science, Bengaluru, 560012, India}
\affil[*]{Email: \href{mailto:saigautamg@iisc.ac.in}{saigautamg@iisc.ac.in}}

\date{}

\begin{document}

\maketitle

\begin{abstract}
We assess the accuracy and computational efficiency of the recently developed meta-generalized gradient approximation (metaGGA) functional, the restored regularized strongly constrained and appropriately normed (r$^2$SCAN), in transition metal oxide (TMO) systems and compare its performance against SCAN. Specifically, we benchmark the r$^2$SCAN-calculated  oxidation enthalpies, lattice parameters, on-site magnetic moments, and band gaps of binary 3\textit{d} TMOs against the SCAN-calculated and experimental values. Additionally, we evaluate the optimal Hubbard \emph{U} correction required for each transition metal (TM) to improve the accuracy of the r$^2$SCAN functional, based on experimental oxidation enthalpies, and verify the transferability of the \emph{U} values by comparing against experimental properties on other TM-containing oxides. Notably, including the \textit{U}-correction to r$^2$SCAN increases the lattice parameters, on-site magnetic moments and band gaps of TMOs, apart from an improved description of the ground state electronic state in narrow band gap TMOs. The r$^2$SCAN and r$^2$SCAN+\textit{U} calculated oxidation enthalpies follow the qualitative trends of SCAN and SCAN+\emph{U}, with r$^2$SCAN and r$^2$SCAN+\textit{U} predicting marginally larger lattice parameters, smaller magnetic moments, and lower band gaps compared to SCAN and SCAN+\textit{U}, respectively. We observe that the overall computational time (i.e., for all ionic+electronic steps) required for r$^2$SCAN(+\textit{U}) to be lower than SCAN(+\textit{U}). Thus, the r$^2$SCAN(+\textit{U}) framework can offer a reasonably accurate description of the ground state properties of TMOs with better computational efficiency than SCAN(+\textit{U}). 
\end{abstract}


\section{Introduction}
Density functional theory (DFT\cite{Kohn1996}) calculations are the bedrock of modern computational materials science in terms of predicting thermodynamic and kinetic properties, with such property predictions being put to use in subsequent materials discovery\cite{Jain2016, zhang2018, canepa2017, oganov2019, kirklin2013, penev2021} and understanding underlying physical phenomena.\cite{Singh2017,Hasnip2014, chan2012, zhou2021, bhattacharya2011}  In recent years, machine learning has been used to augment DFT in property predictions, thereby reducing computational cost and accelerating materials discovery.\cite{Schindler2020, Gong2021, ouyang2019, park2020, duan2021} Note that a key approximation within DFT is the exchange-correlation (XC) functional, the exact form of which is unknown. However, several approximations for the XC functional have been proposed over the years, which can be categorized into different classes depending on the degree of sophistication and accuracy, and visually represented as rungs on the Jacob's ladder.\cite{Kohn1996, Jain2016, Tran2016, Perdew2001} As with most computational tools, the higher the accuracy (higher up Jacob's ladder) higher is the computational cost.

Most DFT calculations for “large” solid systems (10s to 100s of atoms) are performed using the Perdew-Burke-Ernzerhof (PBE) parameterization of the generalized gradient approximation (GGA) XC functional,\cite{Perdew1996} as it offers fair accuracy at reasonable computational cost for a wide variety of materials.\cite{holger2021, jadidi2020, zhang2020} Specifically, GGAs include the local electron density as well as the gradient of the electron density in describing the XC. As a semilocal functional of electron density, PBE captures short range interactions but fails to capture medium and long-range dispersions and also exhibits large electronic self-interaction errors (SIEs), especially in highly correlated systems.\cite{sharkas2020, vargas2020} Also, PBE typically underestimates the formation energies\cite{Isaacs2018, sarmiento2015} and semiconductor band gaps of crystalline solids,\cite{yang2016, Isaacs2018} while overestimating their lattice volumes.\cite{Isaacs2018, perdew2008} 

As we move higher in the Jacob’s ladder,\cite{Perdew2001} we obtain metaGGA functionals, which may account for medium range dispersions and exhibit lower SIEs. Some metaGGAs consider orbital kinetic energy density in addition to the local electron density and its gradient, such as the recently developed strongly constrained and appropriately normed (SCAN\cite{Sun2015}) functional, which offers better numerical accuracy than PBE and satisfies all 17 known constraints for a XC functional (namely, 6 for exchange, 6 for correlation, and 5 for both). The iso-orbital indicator ($\alpha$), which includes the kinetic energy density in SCAN, distinguishes various bonding environments in a given material and consequently improves the accuracy of SCAN over GGA. However, SCAN suffers from numerical instability during self-consistent-field (SCF) calculations\cite{Rodriguez2020} wherein denser $k$-grids (than PBE) are required for accurate and consistent predictions.\cite{Furness2020, Ehlert2021,Rodriguez2020} Thus it is computationally expensive (per SCF step) compared to PBE.\cite{holger2021}

To overcome the numerical instability and reduce the computational cost of SCAN, Bartok and Yates\cite{bartok} developed regularized SCAN (rSCAN), which satisfies 13 out of the 17 known constraints. The authors replaced the non-analytical switching $\alpha$ interpolation function in SCAN with a simple polynomial function, which improves computational speed.\cite{Furness2021} However, subsequent investigations showed a significant drop in numerical accuracy with rSCAN (compared to SCAN), which is attributed to the failure of the polynomial $\alpha$ function to fully recover the uniform gas limit.\cite{Furness2020, Rodriguez2020} Subsequently, Furness et al.\cite{Furness2020} introduced the restored regularized SCAN (or r$^{2}$SCAN), wherein the constraints broken by rSCAN were restored except the fourth order gradient expansion constraint for exchange (or GE4X). Furness et al.\ claimed that the new r$^{2}$SCAN functional combines the numerical accuracy of SCAN and computational speed of rSCAN as the smooth polynomial $\alpha$ function of rSCAN is modified to satisfy the uniform gas limit in r$^2$SCAN.\cite{Furness2020} Recently, Kingsbury et al.\cite{Kingsbury2021} demonstrated that r$^{2}$SCAN functional indeed delivers robust numerical accuracy (i.e., similar to SCAN) and better computational performance (faster and numerically stable) by comparing r$^{2}$SCAN and SCAN for solids using a high-throughput computational workflow. Specifically, the authors\cite{Kingsbury2021} reported that while r$^{2}$SCAN predicts a smaller band gap (for most of the strongly-bound materials) and larger lattice volumes than SCAN, the mean atomization error with r$^{2}$SCAN is $\sim$15-20\% lower for most solids. However, the performance of r$^{2}$SCAN in correlated electron systems, i.e., transition metal oxides (TMOs) containing open-shell \textit{d} electrons, remains to be seen and forms the main focus of this work. 

Despite the accuracy of SCAN, it still has shortcomings in TMOs, which can be mitigated by adding an on-site Hubbard \textit{U} correction term for the transition metal (TM) under consideration.\cite{Gautam2018, Long2020} This approach is similar to the one followed to mitigate the SIEs of PBE in TMOs.\cite{Anisimov1991, Franchini2007} However, the magnitude of the \textit{U} correction required is not known \textit{a priori}, and there are both theory-based approaches such as density functional perturbation theory,\cite{Timrov2018} linear response theory,\cite{Cococcioni2005, Zhou2004, Moore2022} embedded Hartree-Fock method,\cite{Mosey2007, Mosey2008} and machine learning based Bayesian optimisation,\cite{Yu2020} and experimental-data-based approaches to identify the appropriate \textit{U} values. For example, Gautam et al.\cite{Gautam2018, Long2020} used the experimental oxidation enthalpies among binary TMOs to identify optimal \textit{U} values across various oxidation states of 3\textit{d} TMs. A similar experimental-data-based Hubbard \textit{U} correction scheme can be developed in conjunction with r$^{2}$SCAN as well, resulting in a r$^{2}$SCAN+\textit{U} framework, in case r$^{2}$SCAN exhibits similar SIEs as SCAN in TMOs. We explore the usefulness of such a r$^{2}$SCAN+\textit{U} framework also in this work.

Here, we verify the numerical accuracy and computational efficiency of the r$^2$SCAN and r$^2$SCAN+\textit{U} frameworks in comparison to SCAN and SCAN+\textit{U}, respectively, in describing material properties such as lattice parameters, on-site magnetic moments, and band gaps of binary 3\textit{d} TMOs, including Ti, V, Cr, Mn, Fe, Co, Ni, and Cu. As necessary, we evaluate the optimal Hubbard \textit{U} correction with r$^2$SCAN for each TM by using the experimental-data-based approach employed in previous works.\cite{Gautam2018, Long2020} We find that r$^2$SCAN predicts marginally larger lattice constants and smaller on-site magnetic moments than SCAN for most of the TMOs considered. On addition of the \textit{U}-correction to both SCAN and r$^2$SCAN, we observe an increase in the calculated lattice constants, on-site magnetic moments and band gaps. In the case of narrow band gap TMOs, SCAN+\textit{U} and r$^2$SCAN+\textit{U} generally estimate a non-zero band gap, with r$^2$SCAN+\textit{U}'s band gap in better agreement with experiments. Also, we perform transferability checks for the optimal \emph{U} values derived in this work for each TM, by benchmarking various properties in oxides that were not used in obtaining the \emph{U} values. Finally, we compare the computational performance of r$^2$SCAN/r$^2$SCAN+\textit{U} relative to SCAN/SCAN+\textit{U} to explore the accuracy-cost trade-off. We report that r$^2$SCAN/r$^2$SCAN+\textit{U} is computationally less expensive than SCAN and SCAN+\emph{U}, when all required ionic and electronic steps are taken into account for convergence during structure relaxations. We hope that our work will provide a foundational basis for further studies on understanding material behavior and computationally discovering new materials in the near future.

\section{Methods}
\subsection{Computational Methods}
We used the Vienna ab-initio simulation package (VASP 6.2.1)\cite{Hafner1997, kresse1993, kresse1996} for all the spin-polarized DFT calculations, where the frozen-core PBE-based projector augmented wave (PAW)\cite{Kresse} potentials employed were identical to previous work.\cite{Gautam2018, Long2020} The plane waves for each system were expanded up to a kinetic energy of 520~eV, with each structure converged until the total energy differences and atomic forces became \textless 0.01~meV and \textless $|0.01|$~eV/Å, respectively.  We adopted a $\Gamma$-centered Monkhorst-Pack\cite{Monkhorst1976} grid with a density of 48 $k$-points per Å for all systems. The conjugate gradient algorithm was used to relax the structures (i.e., cell shapes, volumes, and ionic positions), without preserving any underlying symmetry. An `accurate' level of precision was maintained while projecting the wavefunctions in the reciprocal space. The Fermi surface of each system was integrated with a Gaussian smearing of partial occupancies, with a width of 0.05~eV. In terms of DFT+\textit{U} calculations, we used the Dudarev framework\cite{Dudarev1998} for adding a effective \textit{U} correction on the $d$ orbitals of TM atoms. All \textit{U} values used in SCAN+\textit{U} calculations were taken from previous work (see Table~S1 of the Supporting Information --SI).\cite{Gautam2018, Long2020} Since we used different computing systems to perform our structure relaxations for different systems, we normalized the computational time with the number of cores used in each calculation to compare the computational efficiency of the different XC functionals considered. 

For calculating band gaps, GGA functionals typically use the Kohn Sham potential as a multiplicative term, which typically underestimates the band gap of solids even at the SCAN level.\cite{Kitchaev2016, Perdew2017} Here, we use the generalized Kohn Sham technique to determine the band gaps by calculating the density of states (DOS) for all systems considered. For each DOS calculation, we used the optimized structure and the initial charge density from a previous structure relaxation. Subsequently, we introduced a set of zero-weighted $k$-points, corresponding to a density of 96 $k$-points per Å, where the $k$-points that were used for the structure relaxation retained their original weights (as determined by VASP). Finally, we performed a single-SCF calculation where the DOS was sampled between energies of -20 to 20 eV in steps of 0.005 eV.


\subsection{Structures and magnetic configurations}
We considered the binary oxides of each TM, i.e., Ti, V, Cr, Mn, Fe, Co, Ni, and Cu with different oxidation states, similar to previous studies.\cite{Gautam2018, Long2020} The main criteria in selection of these metal oxides are the availability of reliable thermodynamic data (i.e., formation energies\cite{kubaschewski1993materials, Wagman, barin}) and the experimentally-determined ground-state structures that are compiled in the inorganic crystal structure database (ICSD)\cite{hellenbrandt2004inorganic} Note that the structures from the ICSD were the initial structures in all our DFT structure relaxations, including the systems used as transferability checks. In the case of Ni oxides, we chose NiO and LiNiO$_2$ (similar to previous work,\cite{Long2020}), as reliable thermodynamic data is not available for higher-oxidation-state binary Ni oxides (e.g., Ni$_2$O$_3$ and NiO$_2$). The TM in all oxides, except select Co and Ni compounds, was initialized in its high-spin configuration (e.g., high-spin configuration of Fe$^{3+}$ consists of five unpaired $d$ electrons). A detailed description of all structures utilised in this work is provided in the SI, under the 'Crystal Structures' section, with the magnetic configurations depicted in Figure~S1.

The magnetic configuration of each TMO considered (see Figure~S1) was initialized to its appropriate (in several cases, experimentally-known) ground state configuration during the structural relaxation. For example, we considered the ferromagnetic (FM) ground state configuration for CrO$_2$ and VO$_2$, given that CrO$_2$ is metallic{\cite{coey2002}} and VO$_2$ undergoes a metal-to-insulator transition (MIT) below 341~K.{\cite{Rogers1993}} The rocksalt (RS) TMOs, namely, VO, MnO, FeO, CoO, and NiO were initialized with their experimentally-known type-II antiferromagnetic (AFM) configuration.{\cite{sasaki1979,kuriyama1962,Jauch2001,Shen1991,Helmer,hartmann1951}} Each Ni's spin in NiO was initialized with two unpaired $d$ electrons (i.e., its high-spin configuration). In CuO, we arranged the magnetic moments of Cu$^{2+}$ antiferromagnetically along the Cu-O-Cu chains in the [10$\bar{1}$] direction.{\cite{Ghijsen1988, mittal2007}}

We initialized $\alpha$-Mn$_2$O$_3$ (bixbyite structure) in a FM configuration as this configuration was found to be the most stable in previous work.{\cite{Gautam2018}} AFM configurations were utilized for rutile-MnO$_2$,{\cite{regulski2003}}, and the other TM$_2$O$_3$ oxides, namely, V$_2$O$_3$, Fe$_2$O$_3$, Ti$_2$O$_3$, and Cr$_2$O$_3$. Note that V$_2$O$_3$ becomes AFM below its MIT temperature,{\cite{Loehman,Rozier2002,Ashkenazi1975}} while Fe$_2$O$_3$ displays an AFM configuration with the magnetic moment of Fe alternating every two consecutive layers along the $c$-axis.{\cite{Maslen1994}} Cr$_2$O$_3$ and Ti$_2$O$_3$ exhibit $\uparrow\downarrow\uparrow\downarrow$ and $\uparrow\downarrow\downarrow\uparrow$ magnetic configurations, respectively, on the TM centers along the $a$-axis.{\cite{brockhouse1953,abrahams1963}} 

In case of spinels, we used different ferrimagnetic (FIM) configurations, as per experimental observations. For example, spinel-Fe$_3$O$_4$ contains both Fe$^{3+}$ and Fe$^{2+}$, with up-spin Fe$^{3+}$ occupying tetrahedral sites and down-spin Fe$^{3+}$ occupying half the octahedral sites. The remaining octahedral sites in Fe$_3$O$_4$ are occupied by up-spin Fe$^{2+}$.{\cite{verwey1939,park1998}} In Co$_3$O$_4$, no-spin Co$^{3+}$ occupies octahedral sites, while high-spin Co$^{2+}$ (three unpaired $d$ electrons) occupies tetrahedral sites in an AFM configuration.{\cite{Smith1973, Knop1968, Roth1964}} For Mn$_3$O$_4$, we adopted the "FIM6" configuration, as this was found to be the ground state in previous work.\cite{Gautam2018} TiO$_2$, CrO$_3$, and V$_2$O$_5$ are diamagnetic, since they contain TMs with empty 3$d$ orbitals. Similarly, Cu$_2$O is diamagnetic owing to the completely-filled 3\textit{d} orbitals of Cu. 

\subsection{Determining $U$}
We determined the required \textit{U} value, with r$^2$SCAN, for each binary TMO oxidation reaction (e.g., Ti$^{3+}\rightarrow$ Ti$^{4+}$ in 2Ti$_2$O$_3$ + O$_2 \rightarrow$ 4TiO$_2$) by comparing the experimental enthalpy (per mole of O$_2$) with the calculated (r$^2$SCAN+\textit{U}) value that minimizes the error against the experimental value. Note that \textit{U} = 0 eV in our data simply reflects a r$^2$SCAN calculation. In order to obtain the experimental oxidation enthalpy, standard enthalpy of formation for all the considered TMOs were taken from the Wagman and/or Kubaschewski tables,\cite{kubaschewski1993materials, Wagman} thus ignoring the $p-V$ and entropic contributions, similar to previous works.\cite{Gautam2018,Long2020,Wang2004} The overall optimal \textit{U} value for each TM was obtained by taking the average of the required \textit{U} for each of the available oxidation reactions. In the case of Ni oxides, oxidation of NiO to LiNiO$_2$ by 2Li$_2$O + 4NiO + O$_2$ $\rightarrow$ 4LiNiO$_2$ was considered as a proxy for the Ni$^{2+} \rightarrow$ Ni$^{3+}$ oxidation reaction.\cite{Long2020} 

\section{Results}
\subsection{Oxidation energetics}
\begin{figure}[h!]
    \centering
    \includegraphics[width=1\textwidth]{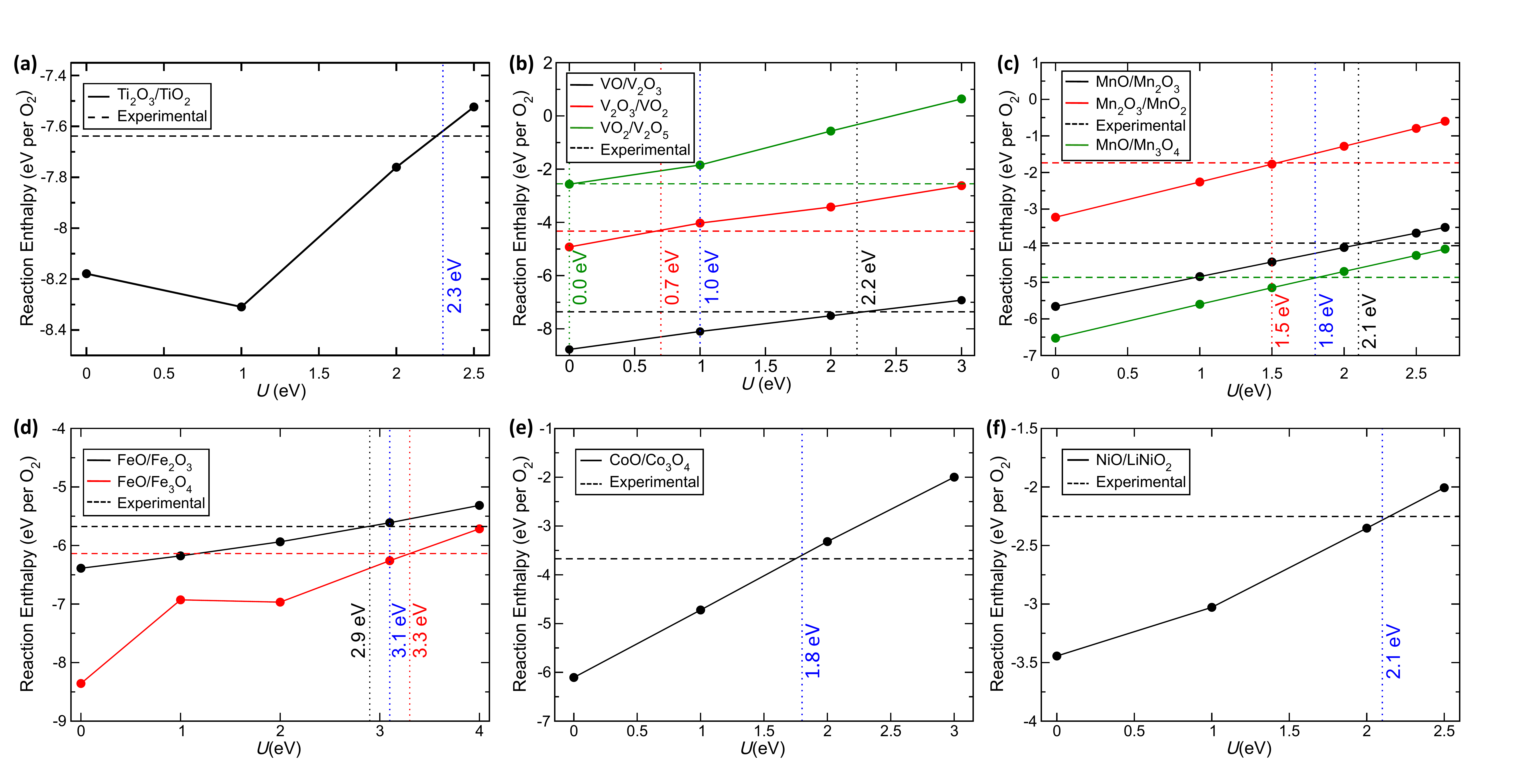}
    \caption{Calculated oxidation enthalpy versus the magnitude of \textit{U} correction within r$^2$SCAN+\textit{U} framework for (a) Ti, (b) V, (c) Mn, (d) Fe, (e) Co, and (f) Ni oxides. Solid, dashed, and dotted lines of a given color indicate calculated, experimental, and required \textit{U} values for a given oxidation reaction. Optimal \textit{U} for each TM is indicated by the dotted blue line in each panel.}
    \label{Oxidation enthalpy vs U}
\end{figure}

Figure~{\ref{Oxidation enthalpy vs U}} displays the variation of the enthalpy of different oxidation reactions among binary TMOs, as a function of applied \textit{U} in the r$^2$SCAN+\textit{U} framework, for all TMs considered except Cr and Cu. Solid lines in each panel of Figure~{\ref{Oxidation enthalpy vs U}} represent DFT-calculated oxidation enthalpies, with each color corresponding to different oxidation reactions for the TM. For instance in V oxides (Figure~{\ref{Oxidation enthalpy vs U}}b), the solid black line corresponds to the oxidation reaction, VO $\rightarrow$ V$_2$O$_3$, while the solid red and green lines indicate V$_2$O$_3 \rightarrow$ VO$_2$ and VO$_2 \rightarrow$ V$_2$O$_5$, respectively. Similarly, the experimental enthalpy of each oxidation reaction is represented by dashed horizontal line of the same color. For example, the black dashed line in Figure~{\ref{Oxidation enthalpy vs U}}b indicates the experimental oxidation enthalpy (-7.36~eV) of VO $\rightarrow$ V$_2$O$_3$. Also, dotted vertical line of a given color highlights the required \textit{U} value to minimize the error between DFT-calculated and experimental value for the oxidation reaction enthalpy indicated by the same color. The dotted blue line in each panel signifies the overall optimal \textit{U} for the TM that is averaged across all available oxidation reactions. 

We report an optimal \textit{U} value of 2.3, 1.0, 1.8, 3.1, 1.8, and 2.1~eV, respectively, for Ti, V, Mn, Fe, Co, and Ni oxides, within the r$^2$SCAN+\textit{U} framework (Figure~{\ref{Oxidation enthalpy vs U}}). Notably, the optimal \textit{U} obtained with r$^{2}$SCAN is less than that reported previously for SCAN functional (Table~S1) for all 3\textit{d} TMs considered (except V and Fe), which can be attributed to better accuracy of r$^{2}$SCAN compared to SCAN, as observed in non-TMOs.\cite{Kingsbury2021} For V oxides, the required \textit{U} value for VO$_2$ $\rightarrow$ V$_2$O$_5$, V$_2$O$_3$ $\rightarrow$ VO$_2$, VO $\rightarrow$ V$_2$O$_3$ is 0.0, 0.7, and 2.2~eV, respectively. Thus, the optimal \textit{U} value for V is 1.0~eV (average of the three required \textit{U} values), which is identical to the \textit{U} correction required with SCAN.\cite{Long2020} The decreasing required \textit{U} with increasing oxidation state of V in V oxides is expected due to the decrease in the strength of exchange interactions among the \textit{d} electrons as oxidation state increases. In the case of Fe, FeO $\rightarrow$ Fe$_2$O$_3$ and FeO $\rightarrow$ Fe$_3$O$_4$ reactions require a \textit{U} of 2.9 and 3.3~eV, respectively, resulting in an optimal \textit{U} of 3.1~eV, which is also identical to the optimal \textit{U} with SCAN.\cite{Gautam2018} Moreover, we obtain the highest optimal \textit{U} of 3.1~eV for Fe, among all TMs considered in this work, which is consistent with the fact that Fe$^{3+}$ has the highest number of unpaired $d$ electrons resulting in the strongest exchange interactions. 

For Ti and Ni, we observe a marginal improvement in the \textit{U}-value for r$^{2}$SCAN when compared to SCAN. Specifically, we obtain an optimal \textit{U} of 2.3~eV and 2.1~eV for Ti and Ni, respectively, versus 2.5 eV for both elements with SCAN. We find an optimal \textit{U} value of 1.8 eV for both Mn (2.7~eV with SCAN) and Co (3.0~eV with SCAN). In Mn-oxides, the required \textit{U} for the oxidation of Mn$_2$O$_3 \rightarrow$ MnO$_2$, and MnO $\rightarrow$ Mn$_2$O$_3$ are 1.5 and 2.1~eV, respectively. The optimal \textit{U} for Mn is transferable to other Mn oxides as well, indicated by the robust agreement between r$^{2}$SCAN+\textit{U}-calculated and experimental oxidation enthalpy for MnO $\rightarrow$ Mn$_3$O$_4$ (green lines in Figure~{\ref{Oxidation enthalpy vs U}}c).  

For Cr and Cu oxides, we obtain reasonable agreement with experimental data without a \textit{U} correction (Figure~S2), similar to our observation with SCAN.\cite{Long2020} In fact, for Cu, introducing \textit{U}-correction worsens the error in the calculated oxidation enthalpy for Cu$_2$O $\rightarrow$ CuO versus experiment, similar to our observation with SCAN(+\textit{U}) as well, which can be attributed to PAW potentials derived at the PBE-level.\cite{Long2020} However, the magnitude of error (versus experiment) is smaller with r$^{2}$SCAN ($\approx$13.1\%) than with SCAN ($\approx$25.7\%). In case of Cr, the oxidation reaction of CrO$_2 \rightarrow$ CrO$_3$ requires \textit{U} $\sim$~0.9 eV, but introducing a \textit{U} correction worsens any agreement with experiment for  Cr$_2$O$_3 \rightarrow$ CrO$_2$ (where required \textit{U}~=~0~eV). Thus, the optimal \textit{U} for Cr oxides is 0.45 eV (\textless 0.5 eV), which only provides a marginal improvement in describing oxidation enthalpies. Hence, we recommend using only r$^{2}$SCAN for calculating any Cr oxide framework.

\subsection{Lattice parameters}
\begin{figure}[h!]
    \centering
    \includegraphics[width=1\textwidth]{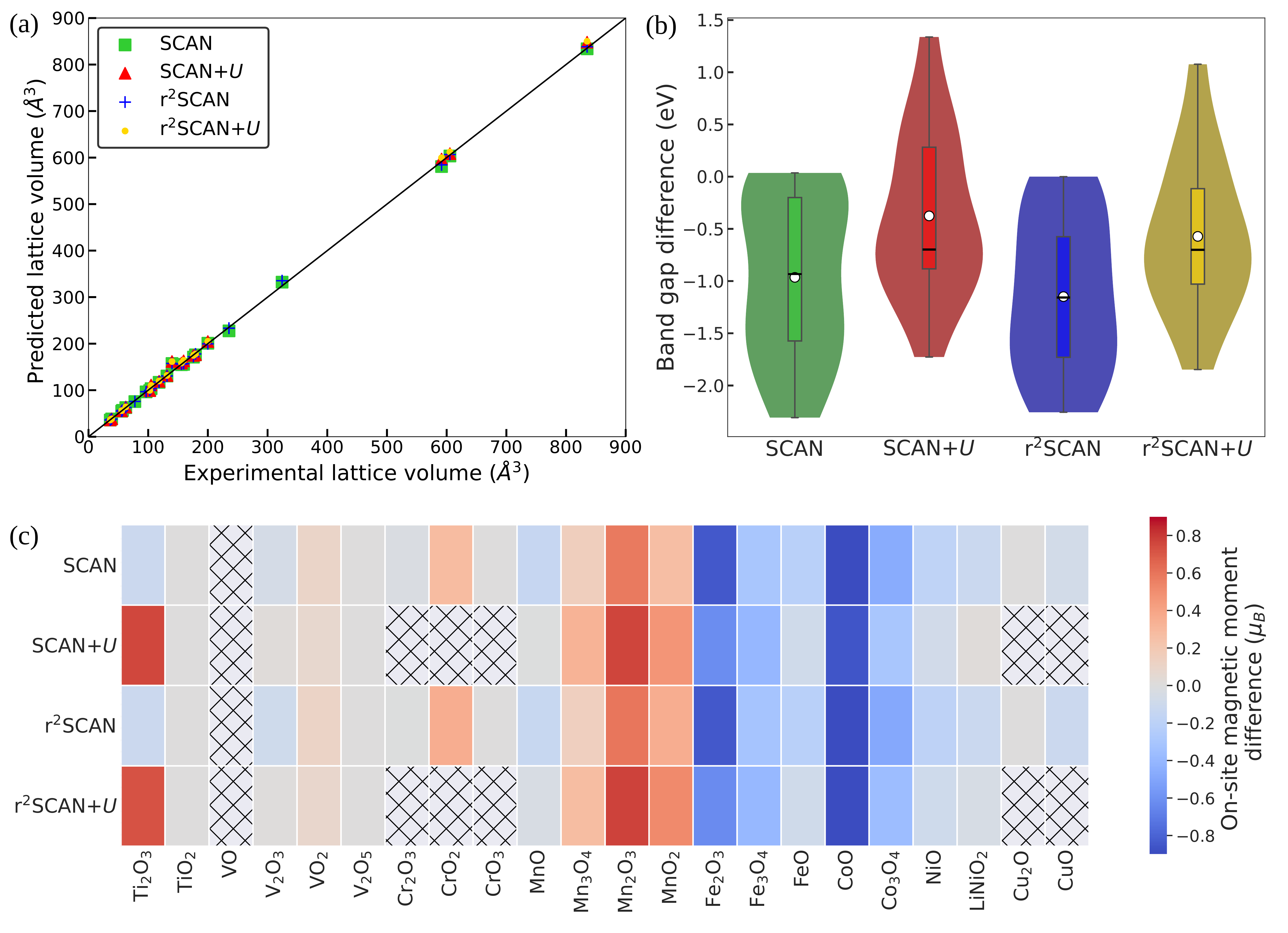}
    \caption{(a) Comparison of calculated and experimental lattice volume (in \AA{}$^3$) of all TMOs considered. (b) Violin plot capturing the difference between the experimental and computed band gap (in eV) across TMO systems using the four XC frameworks. The empty circle and horizontal line in the inner box plot corresponds to the mean and median of the calculated band gaps, respectively. (c) Heat map representation of the differences between the experimental and calculated on-site magnetic moments (in $\mu_B$) using the four XC functionals and across all TMOs. A value of zero indicates perfect consistency, while red (blue) colors indicate overestimation (underestimation) of magnetic moments. Hatched boxes either correspond to experimentally undetermined magnetic moments (VO) or calculations not executed with \emph{U} frameworks (Cr and Cu oxides).}
    \label{Compare}
\end{figure}

All r$^2$SCAN(+\textit{U}) and SCAN(+\textit{U}) calculated lattice parameters, on-site magnetic moments, and band gaps for each TMO are tabulated in Table~S2. Additionally, the calculated lattice volumes by the four XC functionals are plotted against experimental data in Figure~\ref{Compare}a for all oxides. Generally, both SCAN (green squares in Figure~\ref{Compare}a) and r$^2$SCAN (blue symbols) offer $<$~2.8\% deviation from the experimental lattice parameters for all the TMOs considered, except VO, FeO, CuO, and LiNiO$_2$, indicating robust agreement with experiments for both functionals. In VO, SCAN and r$^2$SCAN overestimate (by $\sim$8\%) the experimental lattice constants, while the deviation in FeO and CuO is $\sim$3-4\% and $\sim$8-10\%, respectively. In LiNiO$_2$, SCAN's $\beta$ angle evaluation is $\sim$4.1\% different from experiment. 

Notably, SCAN and r$^2$SCAN do show qualitative differences in their calculated lattice parameters (when compared against experiments) across TMOs. For instance, both functionals overestimate the experimental lattice constants in TiO$_2$, Ti$_2$O$_3$, and VO, while they underestimate in CrO$_2$, CrO$_3$, MnO$_2$, and Fe$_3$O$_4$. There are also examples (MnO and Mn$_2$O$_3$) where SCAN underestimates the experimental lattice constants while r$^2$SCAN overestimates. Overall, there are cases where SCAN's errors in lattice parameter estimations are lower versus experiments (e.g., Cr$_2$O$_3$, CoO), r$^2$SCAN's errors are lower (e.g., CrO$_2$, CrO$_3$, MnO$_2$, Fe$_3$O$_4$), and both functionals exhibit identical errors (e.g., TiO$_2$, Co$_3$O$_4$, NiO, Cu$_2$O), signifying that both functionals offer similar performance in terms of geometrical properties. 

Comparing r$^{2}$SCAN and SCAN, we find that r$^2$SCAN's lattice constants are generally larger than SCAN across TMOs (e.g., Ti$_2$O$_3$, Cr$_2$O$_3$, CrO$_3$, VO$_2$, etc.). As a range, r$^2$SCAN estimates lattice constants that are a maximum of $\sim$1.5\% larger than SCAN (in CrO$_3$) and a minimum of $\sim$0.1\% larger than SCAN (in Mn$_2$O$_3$). Having said that, there are instances where r$^2$SCAN's lattice constant evaluations are lower than SCAN (VO, CoO, CuO, and LiNiO$_2$) and cases where both functionals are identical (TiO$_2$, Co$_3$O$_4$, NiO, and Cu$_2$O). In specific TMOs, SCAN and r$^2$SCAN calculate an identical (individual) lattice constant, while the other lattice constants with r$^2$SCAN are larger than SCAN. For example, $a$ and $c$ lattice constants with r$^2$SCAN are higher than SCAN in V$_2$O$_5$ while both functionals estimate $b$~=~3.55~\AA. 

On introducing the optimal \textit{U} correction, an increase in the value of calculated lattice constants is obtained for both SCAN and r$^2$SCAN functionals for all TMOs. The lattice constants computed by r$^2$SCAN+\textit{U} (yellow symbols in Figure~{\ref{Compare}}a) is up to 1.3\% higher than r$^2$SCAN, except FeO ($\sim$4.2\% higher). Similar to the comparison of r$^2$SCAN vs. SCAN, there are systems where r$^2$SCAN+\textit{U} predicts larger, smaller, and identical lattice constants compared to SCAN+\textit{U} (red triangles). For example, r$^2$SCAN+\textit{U} calculates larger lattice constants than SCAN+\textit{U} in VO$_2$, V$_2$O$_5$, MnO, Mn$_2$O$_3$ and Fe$_3$O$_4$ (maximum of $\sim$0.5\% higher in V$_2$O$_5$), while for Ti$_2$O$_3$, CoO and NiO, r$^2$SCAN+\textit{U}'s estimations are smaller than SCAN+\textit{U} (maximum deviation of $\sim$2.1\% in Ti$_2$O$_3$). Both SCAN+\textit{U} and r$^2$SCAN+\textit{U} functionals evaluate identical lattice parameters for TiO$_2$, Co$_3$O$_4$ and LiNiO$_2$. 

Overall, lattice constants calculated by SCAN+\textit{U} and r$^2$SCAN+\textit{U} deviate $< \sim$3.3\% from experiments for all TMOs, except VO and VO$_2$ where deviations of $\sim$8.5\% and $\sim$4.6\% are observed, respectively. Adding \textit{U} improves the agreement with experiment for both SCAN and r$^2$SCAN in Co$_3$O$_4$, while r$^2$SCAN+\textit{U} gives the best estimate of the lattice parameters in FeO ($< 1$\% deviation vs.\ experiments) compared to SCAN, SCAN+\textit{U} and r$^2$SCAN. Notably, all functionals break the rocksalt symmetry of VO, MnO, and FeO, while the cubic symmetry of Fe$_3$O$_4$ is retained only by SCAN. In Ti$_2$O$_3$, the hexagonal symmetry is broken by SCAN but the symmetry is preserved by the other frameworks. In summary, we find that the differences in lattice parameter estimations to be minimal across the four functionals on average, with notable exceptions of a few systems. 

\subsection{On-site magnetic moments}
On-site magnetic moments of the TMOs (Figure~\ref{Compare}c and Table~S2) computed by SCAN and r$^2$SCAN generally underestimate experimental values, with the exception of MnO$_2$, Mn$_2$O$_3$, CrO$_2$, and VO$_2$. Note that larger magnetic moments typically indicate stronger localization of \textit{d} electrons. Comparing r$^2$SCAN and SCAN calculations, we find that r$^2$SCAN typically estimates smaller magnetic moments than SCAN but with several exceptions, such as MnO, MnO$_2$, Mn$_2$O$_3$, Cr$_2$O$_3$, and VO$_2$. Thus, on average, SCAN's magnetic moment predictions are in better agreement with experiments. However, in terms of magnitude, moments predicted by r$^2$SCAN deviate by $<$~3\% from SCAN's estimates, except CuO ($\sim$~6.8\% deviation), CrO$_2$ ($\sim 3.5$\%), and MnO$_2$ ($\sim3.5$\%), highlighting that the differences in the predictions are marginal.

Adding optimal \textit{U} to both SCAN and r$^2$SCAN increases the magnitude of the calculated on-site magnetic moments for all TMOs (except VO$_2$, which is predicted to be metallic by all functionals), consistent with the expectation that the \textit{U} correction facilitates \textit{d} electron localization. r$^2$SCAN+\textit{U}-calculated data are similar to the corresponding SCAN+\textit{U} values ($< 2.3$\% variation), except LiNiO$_2$ ($\sim$6.3\% variation), and Ti$_2$O$_3$ ($\sim$3.8\%). Similar to r$^2$SCAN versus SCAN, r$^2$SCAN+\textit{U} estimates smaller magnetic moments than SCAN+\textit{U}, with notable exceptions being VO$_2$, Mn$_2$O$_3$, MnO$_2$ and FeO. Overall, we observe the accuracy in calculated on-site magnetic moments versus experiments to follow the order SCAN+\textit{U}$>$ r$^2$SCAN+\textit{U} $>$ SCAN $>$ r$^2$SCAN for several TMOs. However, there are specific cases where specific XC frameworks offer better accuracy in calculating magnetic moments, such as SCAN in CrO$_2$, Mn$_2$O$_3$, MnO$_2$, Fe$_3$O$_4$ and CuO, r$^2$SCAN in Mn$_3$O$_4$ and Cr$_2$O$_3$, and r$^2$SCAN+\textit{U} in V$_2$O$_3$. Given the numerically marginal deviations in calculated magnetic moments across the XC frameworks ($\sim$10\% deviation), we expect an increase/decrease in accuracy to be marginal amongst the XC frameworks considered.

\subsection{Band gaps}
The differences between calculated and experimental band gaps of all TMOs considered are visualized as violin plots for SCAN (green violin), SCAN+$U$ (red), r$^2$SCAN (blue), and r$^2$SCAN+$U$ in Figure~{\ref{Compare}}b. The top and bottom ends of the individual violins mark the highest and lowest differences in the respective calculated data. Note that the mean values (white empty circles) are similar for SCAN and r$^2$SCAN, and in turn are lower than their \textit{U}-corrected versions. In other words, addition of the $U$-correction reduces the error of calculated band gaps compared to experimental values, which is expected given that semi-local DFT typically underestimates band gaps. Also, we find that SCAN+$U$ displays the lowest mean band gap difference among the XC functionals considered, indicating that on-average SCAN+$U$ provides better computed band gaps.

\begin{figure}[h!]
    \centering
    \includegraphics[width=1\textwidth]{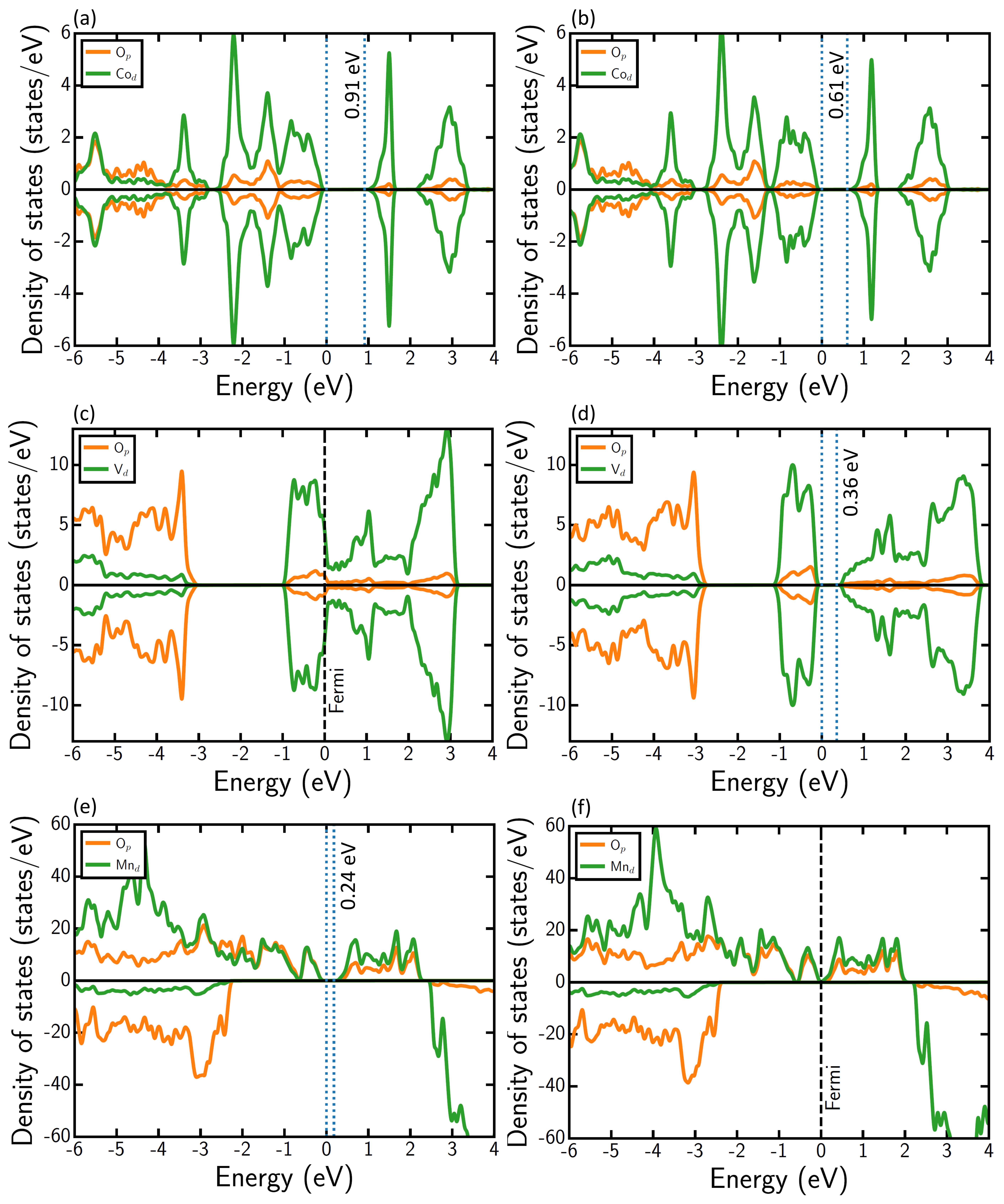}
    \caption{DOS for CoO calculated using (a) SCAN and (b) r$^{2}$SCAN, DOS for V$_2$O$_3$ computed using (c) r$^{2}$SCAN and (d) r$^{2}$SCAN+\textit{U}, and DOS for Mn$_2$O$_3$ estimated using (e) SCAN+\textit{U} and (f) r$^{2}$SCAN+\textit{U}.}
    \label{DOS}
\end{figure}

We present calculated electronic DOS of select TMOs, namely CoO (panels a and b), V$_2$O$_3$ (c and d), and Mn$_2$O$_3$ (e and f), in Figure~{\ref{DOS}}, to illustrate qualitative trends in computed band gaps. The DOS for the remaining TMOs, calculated by the four XC frameworks, are compiled in Figures~S3-S19 of the SI. In each DOS panel, solid orange and solid green lines correspond to the 2\textit{p}-states of O and the 3\textit{d}-states of the TM, respectively. Dashed black lines represent Fermi levels in metallic compounds. Dotted vertical lines represent valence and conduction band edges in semiconducting/insulating compounds, with the band gaps indicated by the text annotation near the conduction band minimum (CBM). The zero of the energy scale is set to the valence band maximum (VBM) for TMOs with a band gap and to the Fermi level in metallic TMOs. 


We observe that r$^2$SCAN generally calculates a smaller band gap than SCAN for most TMOs (maximum of $\sim$66\% lower in MnO$_2$, see Table~S2), as illustrated by the case of CoO in panels a and b of Figure~{\ref{DOS}}. Notable exceptions do exist to this observation, such as V$_2$O$_5$ ($\sim$1.7\% larger), CrO$_3$ ($\sim$3.2\%), MnO ($\sim$4.3\%), and Fe$_2$O$_3$ ($\sim$1.7\%), where r$^2$SCAN calculated band gaps are marginally larger than SCAN. Both SCAN and r$^2$SCAN incorrectly describe the ground state electronic configuration of narrow band gap TMOs (i.e., experimental band gaps $< 1$~eV), including  Ti$_2$O$_3$ (Figure~S4), V$_2$O$_3$(Figure~\ref{DOS}c and S3c), VO$_2$ (Figure~S7) and Fe$_3$O$_4$ (Figure~S15) to be metallic, with the exception of MnO$_2$ where both SCAN and r$^2$SCAN estimate a narrow gap (Figures~S12a and S12c). Additionally, both functionals also calculate the wrong electronic structure in the case of a non-narrow-gap semiconductor, Mn$_2$O$_3$ (Figure~S3), which exhibits an experimental gap of 1.2-1.3~eV.\cite{rahaman2015, feng2012} However, SCAN and r$^2$SCAN qualitatively describe the right electronic structure in the case of wide band gap TMOs such as FeO (Figure~S13), Fe$_2$O$_3$ (Figure~S14), and NiO (Figure~S17), with a significant quantitative underestimation of the experimental gaps. In any case, the differences in electronic structure predictions between SCAN and r$^2$SCAN in TMOs are minimal, with SCAN being marginally better in accuracy.

Introducing a \textit{U} correction to SCAN and r$^2$SCAN widens or opens the band gap, especially in narrow band gap TMOs, as illustrated by the case of V$_2$O$_3$ (panels c and d in Figure~{\ref{DOS}}). The opening of band gap with \textit{U} correction is expected since localization of \textit{d} electrons, which form the VBM and/or CBM in 3\textit{d}-TMOs, is faciliated with \textit{U} addition, in turn resulting in a larger gap. However, in the case of VO$_2$ (Figure~S7), adding \textit{U} does not capture the MIT that occurs at low temperatures ($< 341$~K\cite{Rogers1993}) with either SCAN or r$^2$SCAN, causing the erroneous prediction of metallic behavior. Generally, SCAN+\textit{U} calculates a larger band gap than r$^2$SCAN+\textit{U} (Table~S2), as highlighted by the case of Mn$_2$O$_3$ (panels e and f in Figure~{\ref{DOS}}). In fact, SCAN+\textit{U} is the only framework (among those considered) to estimate a band gap in Mn$_2$O$_3$, which is consistent with experiment. Moreover, SCAN+\textit{U}'s evaluations of larger band gaps results in better (poorer) quantitative agreement with experiments in wide (narrow) gap materials, such as MnO and FeO (V$_2$O$_3$ and MnO$_2$). 

Note that SCAN+\textit{U} and r$^2$SCAN+\textit{U} do underestimate the experimental band gaps, similar to SCAN and r$^2$SCAN, in wide gap TMOs. The only exception to this observation is CoO, where SCAN+\textit{U} overestimates the band gap versus experiment (Figure~S3a and Table~S2), as also observed in our previous work.{\cite{Long2020}} In select TMOs, including Fe$_2$O$_3$ and V$_2$O$_5$, r$^2$SCAN+\textit{U}'s band gap is larger than SCAN+\textit{U}, but the magnitude of difference ($\leq 0.2$~eV) is meagre. Thus, for electronic structure predictions, we expect SCAN+\textit{U} to provide the best qualitative and quantitative band gaps across TMOs, among the functionals considered here, especially for wide gap semiconductors/insulators. However, the qualitative trends provided by r$^2$SCAN+\textit{U} are quite robust as well and in small gap semiconductors ($< 1$~eV gap), r$^2$SCAN+\textit{U}'s quantitative accuracy is often better than SCAN+\textit{U}.

\subsection{Transferability checks}
\begin{figure}[h!]
    \centering
    \includegraphics[width=1\textwidth]{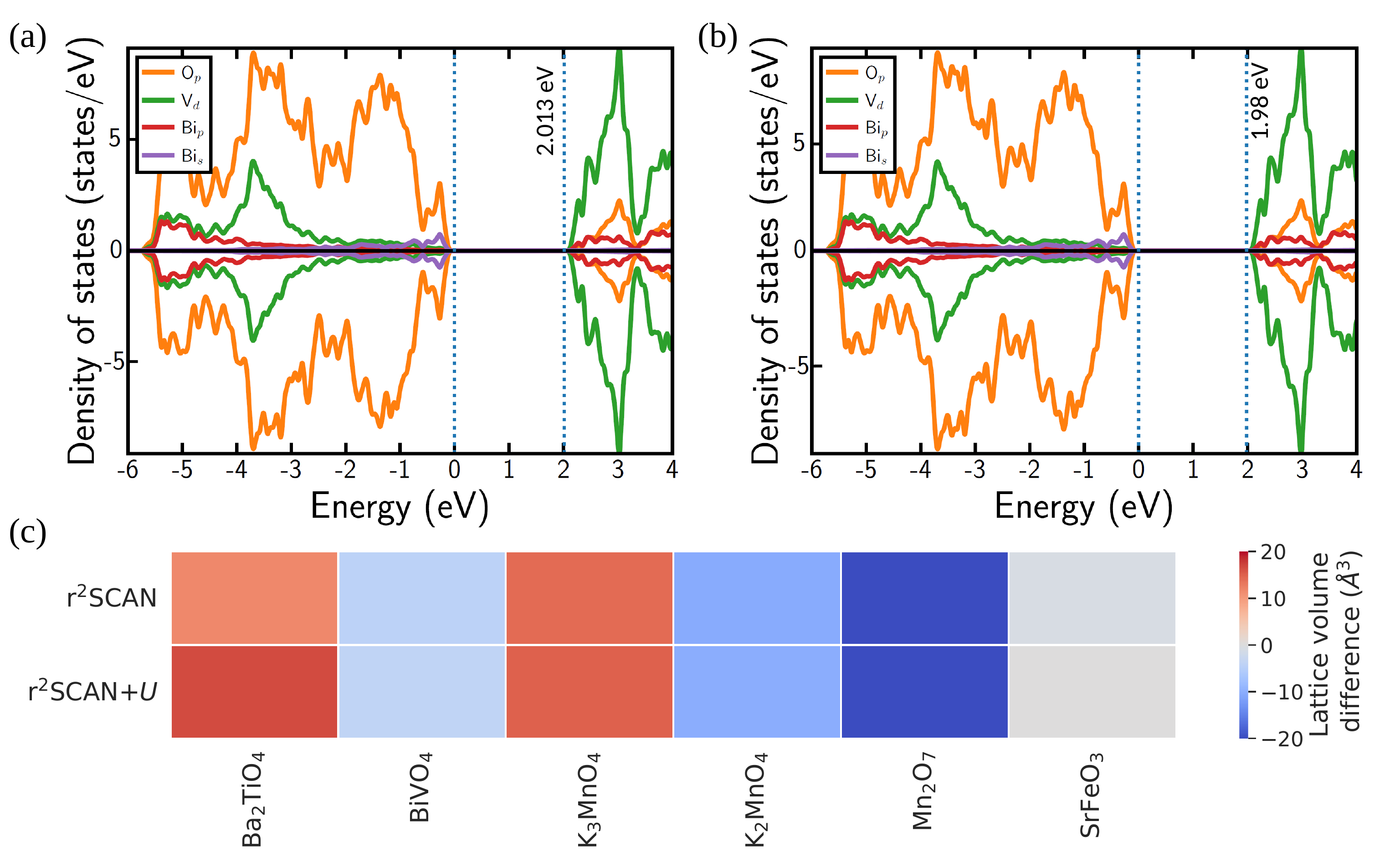}
    \caption{DOS for BiVO$_4$ calculated using (a) r$^{2}$SCAN and (b) r$^{2}$SCAN+\textit{U}. (c) Difference between experimental and calculated lattice volumes (using r$^2$SCAN and r$^2$SCAN+\emph{U}), plotted as a heatmap, for various systems. Red (blue) squares indicate overestimated (underestimated) calculated lattice volumes versus experiment.}
    \label{trns_chk}
\end{figure}

To examine the transferability of the optimal \emph{U} values determined in this work (with r$^2$SCAN), to oxide systems not used for obtaining the values, we perform checks on systems with different oxidation state and/or coordination environment for each TM. We compare calculated values against available experimental data, such as structural, electronic, magnetic, and/or electrochemical properties. Specifically, we choose Ba$_2$TiO$_4$ as a check for Ti, BiVO$_4$ for V, K$_3$MnO$_4$, K$_2$MnO$_4$, and Mn$_2$O$_7$ for Mn, SrFeO$_3$ for Fe, LiCoO$_2$-CoO$_2$ for Co, and LiNiO$_2$-NiO$_2$ for Ni. Data related to transferability checks are compiled in Figure~{\ref{trns_chk}}, Table~{\ref{trf_chk}}, and Table~S3.   

In the case of Ba$_2$TiO$_4$, we compare the calculated lattice parameters with experimental values (see Table~S3 and lattice voliume differences plotted in Figure~{\ref{trns_chk}}). Ba$_2$TiO$_4$ crystallizes in a monoclinic structure (space group \textit{P}2$_1$/\textit{n}) at low temperatures, where the unit cell is composed of four formula units.\cite{bland1961, shanker2004} Ti atoms are present in distorted tetrahedra composed of neighbouring oxygen atoms (TiO$_4$) within the Ba$_2$TiO$_4$ lattice, which is different from the octahedral environments sampled in TiO$_2$ and Ti$_2$O$_3$. Upon structure relaxation, we observe that both r$^2$2SCAN and r$^2$2SCAN+\textit{U} functionals marginally overestimate (by $\sim$2\%) experimental lattice parameters (Figure ~\ref{trns_chk} and Table~S3). Similar to trends observed in Table~S2, adding \emph{U} to r$^2$SCAN increases the calculated lattice parameters in Ba$_2$TiO$_4$ (by $\sim$0.03~\AA{}), thereby marginally reducing the agreement with experiment. 

We benchmark both structural and electronic properties of BiVO$_4$ as a transferability check for V-based systems. Note that BiVO$_4$ transforms from tetragonal (\textit{I}41/\textit{a}) to a monoclinic (\textit{I}2/\textit{b}) `scheelite' phase below $\sim$~528~K,\cite{liu2022, bhattacharya1997} which is a reversible second order ferroelastic transition driven by soft optical phonon modes. The BiVO$_4$ unit cell possesses four formula units, with tetrahedrally coordinated V ions, which is different from the coordination environments of V in VO, V$_2$O$_3$, VO$_2$, and V$_2$O$_5$. Importantly, monoclinic-BiVO$_4$ spontaneously transforms to the tetragonal structure upon structure relaxation with r$^2$2SCAN and r$^2$2SCAN+\textit{U}, similar to the observation by Liu et al \cite{liu2022} with GGA and hybrid functionals. Thus, neither r$^2$2SCAN nor r$^2$2SCAN+\textit{U} predict the correct ground state structure. Additionally, BiVO$_4$ possess a band gap of 2.4–2.48~eV\cite{cooper2014} and is a candidate photocatalyst.\cite{liu2022} Both r$^2$2SCAN and r$^2$SCAN+\emph{U} provide similar band gap predictions (2.01-1.98~eV), which is in good qualitative agreement with experiment. Surprisingly, r$^2$SCAN+\emph{U} evaluates a marginally lower band gap than r$^2$SCAN (see panels a and b in Figure~\ref{trns_chk}). However, both r$^2$2SCAN and r$^2$SCAN+\emph{U} predict similar states occupying the valence band (O$_p$) and conduction band (V$_d$) edges.

The rationale behind the choice of K$_3$MnO$_4$, K$_2$MnO$_4$, and Mn$_2$O$_7$ as checks for Mn-based systems is to explore the higher, unsampled oxidation states of Mn, namely +5, +6, and +7 in K$_3$MnO$_4$, K$_2$MnO$_4$, and Mn$_2$O$_7$, respectively. Also, Mn resides in tetrahedral coordination in these compounds, which is different from the octahedral coordination observed in MnO, Mn$_2$O$_3$, and MnO$_2$. Although Mn$^{2+}$ resides in tetrahedral sites in spinel-Mn$_3$O$_4$, we had not used in the spinel structure to obtain our optimal \emph{U}. We benchmark the calculated lattice parameters versus experiments for all Mn-based transferability checks.

Mn$_2$O$_7$ is a volatile liquid at 298~K and solidifies to a monoclinic crystal structure (\textit{P}2$_1$/\textit{c}) below $\sim$~279~K, with the unit cell consisting of 8 formula units of corner sharing tetrahedral MnO$_4$ pairs.\cite{lawler2017, simon1987} Upon structural relaxation, both r$^2$SCAN and r$^2$SCAN+\textit{U} underestimate the lattice constants of monoclinic-Mn$_2$O$_7$ by $\sim$1-3\% (Figure ~\ref{trns_chk} and Table~S3). In the case of K$_3$MnO$_4$, the tetragonal symmetry (\textit{I}$\overline{4}$2\textit{m}){\cite{olazcuaga1975}} is broken with r$^2$SCAN functional resulting in an orthorhombic structure, while the symmetry is preserved by r$^2$SCAN+\textit{U} (see Figure ~\ref{trns_chk} and Table~S3). Nonetheless, both r$^2$SCAN and r$^2$SCAN+\textit{U} significantly underestimate the $c$ parameter (by $\sim$~13.5\%) and overestimate the $a$ or $b$ parameter ($\sim$~10.2\%). K$_2$MnO$_4$ is an orthorhombic crystal (\textit{Pnma}) with four formula units per unit cell.\cite{palenik1967} Here, r$^2$SCAN and r$^2$SCAN+\textit{U} predict identical lattice parameters, which marginally underestimate experimental values (by $\sim$~0.4-1\%, see Figure ~\ref{trns_chk} and Table~S3).

The choice of SrFeO$_3$, a cubic perovskite, as a check for Fe is largely motivated by the 4+ oxidation state exhibited by Fe in the structure, which is not sampled in FeO, Fe$_2$O$_3$, or Fe$_3$O$_4$. Both r$^2$SCAN and r$^2$SCAN+\emph{U} preserve the cubic symmetry during structure relaxation, with r$^2$SCAN+\emph{U}'s lattice parameters identical to experiments and r$^2$SCAN's parameters being a slight underestimation ($\sim$~0.5\%, see Figure ~\ref{trns_chk} and Table~S3). In terms of magnetic configuration of Fe in SrFeO$_3$, Takeda et al.\cite{takeda1972} reported a helical spin structure via their neutron diffraction experiments, with competing FM and AFM interactions. However, Shein et al.\cite{shein2005} found a FM metallic state to be the ground state of SrFeO$_3$, over a wide range of pressures, based on their first principles calculations, which they attributed to stronger FM than AFM interactions. We considered a FM configuration of Fe atoms in the SrFeO$_3$ unit cell, and the on-site magnetic moments on Fe calculated by both r$^2$SCAN (3.375~$\mu_B$, Table~{\ref{trf_chk}}) and r$^2$SCAN+\textit{U} (3.819~$\mu_B$) overestimate the experimental value (2.7$\pm$0.4~$\mu_B$\cite{takeda1972}). However, our calculated magnetic moments do indicate a localization of $\sim$4 electrons on the $d$ orbitals of Fe, consistent with its +4 oxidation state.

We choose CoO$_2$ (\textit{R}$\overline{3}$\textit{m} or `O3` polymorph\cite{amatucci1996}), and NiO$_2$ (\textit{P}1\textit{m}1 or `O1'\cite{delmas1999}), both layered structures, as transferability checks for Co and Ni, respectively, owing to the unsampled 4+ oxidation states of each TM. In terms of experimental property to benchmark, we choose the average Li intercalation voltage in these structures, i.e., LiCoO$_2$-CoO$_2$, and LiNiO$_2$-NiO$_2$ pairs, since they have been measured with high precision. The reader is referred to previous works on calculating and benchmarking average `topotactic' intercalation voltages.\cite{Aydinol1997,Long2021} r$^2$SCAN underestimates the experimental average voltage\cite{amatucci1996,Delmas1991_Review,Ohzuku1994,Aydinol1997_JPS,Arroyo2002,Long2021} in LiNiO$_2$-NiO$_2$ (by $\sim$~8\%), while it overestimates the average voltage in LiCoO$_2$-CoO$_2$ (by  $\sim$~1.7\%), similar to trends observed with SCAN.\cite{Long2021} The addition of \emph{U} to r$^2$SCAN leads to an improvement in agreement with the experimental voltage in the Ni-system (deviation of $\sim$~1.8\%), while it worsens the agreement in the Co-system (deviation of $\sim$~4.4\%). Nevertheless, r$^2$SCAN+\emph{U} does overestimate the average voltage in both Co and Ni systems, similar to the behavior of SCAN+\emph{U}.\cite{Long2021}

\begin{table}[h!]
    \caption{Voltage and magnetic moments calculated by r$^2$SCAN, and r$^2$SCAN+\textit{U} compared against experimental values (denoted by `Expt.'). The \textit{U} values used with r$^2$SCAN+\textit{U} are the corresponding optimal \textit{U} values obtained for each TM (from Figure~{\ref{Oxidation enthalpy vs U}}).}    
    \centering
    \begin{tabular}{cccccccc}
            \hline
            Composition & Source &  {Voltage}&{Magnetic moment}
            \\
            (space group) & & (V)  & ($\mu_B$) \\\\
            \hline
            LiCoO$_2$-CoO$_2$&Expt.& 4.05 & - \\
            (\textit{R}$\bar{3}$\textit{m})&r$^2$SCAN& 4.12 & -  \\
            &r$^2$SCAN+\textit{U}& 4.23 & -\\
            
            LiNiO$_2$-NiO$_2$&Expt.& 3.85 & - \\
            (\textit{P}1\textit{m}1)&r$^2$SCAN& 3.54 & - \\
            &r$^2$SCAN+\textit{U}& 3.92 & -\\
            
            SrFeO$_3$&Expt.& - &  2.7$\pm$0.4 \\
            (\textit{Pm}$\bar{3}$\textit{m})&r$^2$SCAN& - & 3.375 \\
            &r$^2$SCAN+\textit{U}& - & 3.819\\
            
            \hline
            \end{tabular}

    \label{trf_chk}
\end{table}

\section{Discussion}
In this work, we evaluated the performance of the r$^2$SCAN functional among binary TMOs consisting of 3\textit{d}-TMs by calculating the oxidation enthalpies, lattice parameters, on-site magnetic moments, and band gaps. Additionally, for each TM-O$_2$ system considered, we calculated the optimal Hubbard-\textit{U} corrections to be used in a r$^2$SCAN+\emph{U} framework, based on experimental oxidation enthalpies. Although theoretical approaches exist to derive \textit{U} values,\cite{Mosey2007, Mosey2008, Cococcioni2005, Moore2022, Zhou2004, Timrov2018, Yu2020} using oxidation enthalpies nominally gives an ``average" correction that is suitable across several oxidation states of a given TM. Specifically, our optimal \textit{U} values are 2.3, 1.0, 1.8, 3.1, 1.8, and 2.1~eV for Ti, V, Mn, Fe, Co, and Ni, respectively, while we don't deem a \textit{U} correction necessary for Cr and Cu oxides. Interestingly, the optimal \textit{U} corrections needed with r$^2$SCAN are lower in magnitude compared to SCAN for Ti, Mn, Co, and Ni oxides (while the corrections are identical for V and Fe oxides), indicating that r$^2$SCAN exhibits lower errors with oxidation enthalpies and possibly lower SIEs than SCAN. However, this is not reflected in other physical properties. On an average, we find the accuracy, versus experimental values, to be similar for r$^2$SCAN compared to SCAN, and for r$^2$SCAN+\textit{U} compared to SCAN+\textit{U}, respectively, in lattice parameter, on-site magnetic moment, and band gap evaluations as seen in Figure~{\ref{Compare}}.

\begin{figure}[h!]
    \centering
    \includegraphics[width=0.8\textwidth]{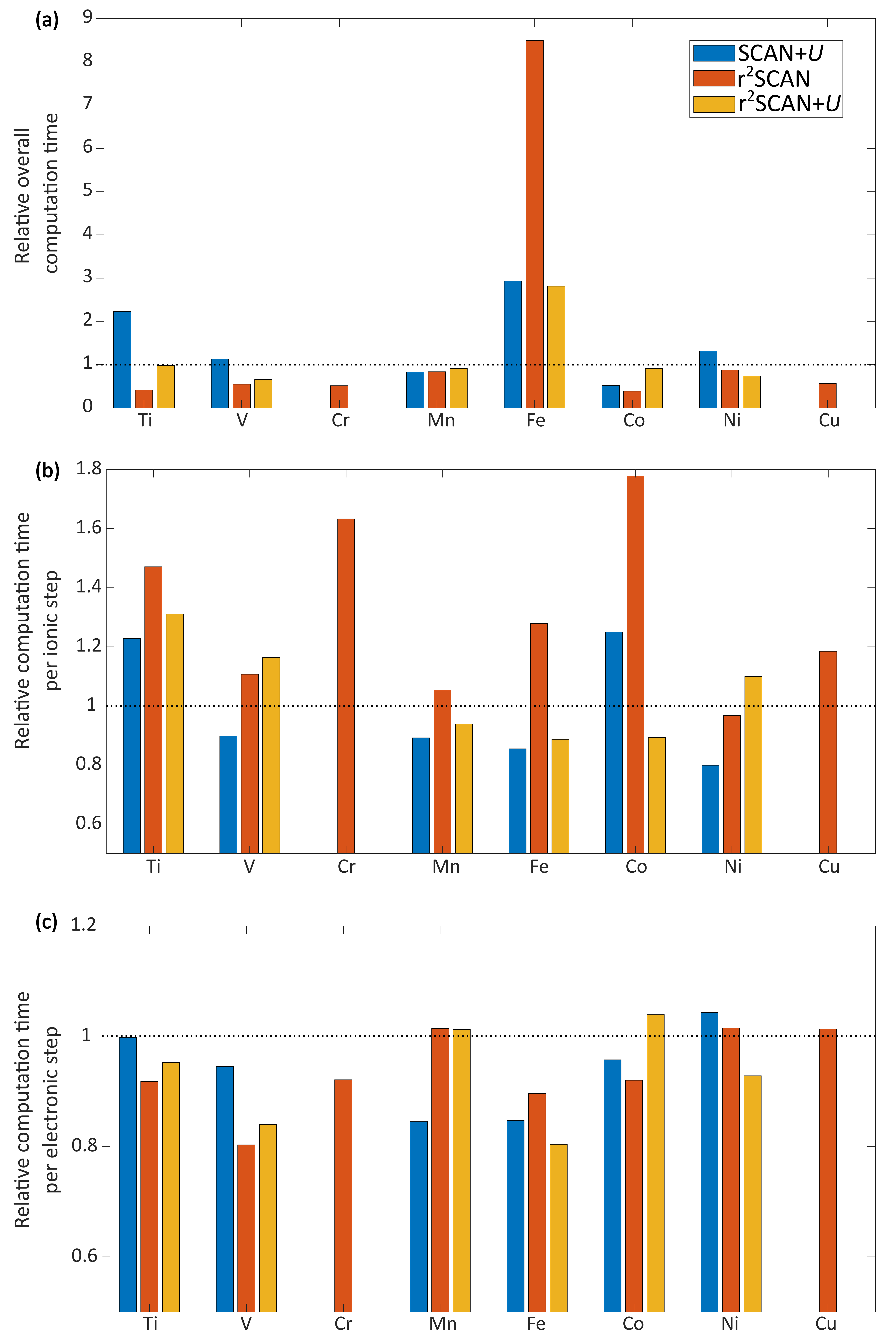}
    \caption{(a) Overall computational time (electronic+ionic steps) (b) computational time per ionic step and (c) computational time per electronic loop taken for each TM-O$_2$ binary system with SCAN+\textit{U}, r$^2$SCAN, and r$^2$SCAN+\textit{U} frameworks relative to SCAN. Values greater (smaller) than 1 in each panel indicates that a given calculation is slower (faster) than SCAN.}
    \label{time}
\end{figure} 

The general trends in lattice parameter, magnetic moment, and band gap predictions, across the XC frameworks considered, can be summarized as follows. We observe that r$^2$SCAN generates larger lattice constants than SCAN and on addition of the \textit{U} correction to both functionals, the lattice constants further increase. Thus, in systems where SCAN underestimates experimental lattice constants (e.g., CrO$_2$, CrO$_3$, MnO$_2$), shifting to r$^2$SCAN improves agreement (e.g., error in r$^2$SCAN in CrO$_3$ is 0.8\% versus 2.3\% with SCAN). Also, there are instances where the ground state symmetry of the TMO is not preserved by some or all of the XC frameworks considered (i.e., in VO, MnO, FeO, Fe$_3$O$_4$, and Ti$_2$O$_3$), highlighting systematic issues in the XC treatment across the four frameworks considered. The calculated on-site magnetic moments by r$^2$SCAN (and r$^2$SCAN+\textit{U}) are marginally lower than SCAN (SCAN+\textit{U}), with the \textit{U} correction nominally increasing the calculated moments calculated by r$^2$SCAN and SCAN. However, calculated magnetic moments across the four XC frameworks differ by $< 10$\% (except LiNiO$_2$), signifying marginal differences in accuracy. Both SCAN and r$^2$SCAN underestimate band gaps across all TMOs (except MnO$_2$), with band gaps calculated by r$^2$SCAN typically being lower than SCAN, and adding the \textit{U} opens/widens the gap. Thus, SCAN+\textit{U} offers the best quantitative accuracy versus experimental band gaps, especially for wide gap semiconductors. Note that the qualitative trends from r$^2$SCAN+\textit{U} are consistent with the trends exhibited by SCAN+\textit{U} and should be reliable in electronic structure predictions in other TM-based oxide systems.  

r$^2$SCAN adopts the smooth polynomial interpolation function of rSCAN to maintain numerical stability during SCF calculations. Additionally, the reformed gradient expansion for correlation introduced in r$^2$SCAN (partially) negates the error introduced to the slowly varying  density by the non-vanishing interpolation function,\cite{Furness2020} which largely accounts for the observed variation in accuracy of r$^2$SCAN versus SCAN. Based on our data, we observe that r$^2$SCAN is not systematically more accurate than SCAN across all TMOs and for all property predictions. For example, we have lower optimal \textit{U} values indicating lower SIEs with r$^2$SCAN versus SCAN, but also lower on-site magnetic moments (except Mn and Cr oxides) signifying poorer \textit{d}-electron localization with r$^2$SCAN. Further, the smaller band gaps with r$^2$SCAN (versus SCAN) may be caused by the residual SIEs, resulting in an underestimation of the CBM across TMOs. Hence, usage of r$^2$SCAN(+\textit{U}) in TM-based systems must be done with care and efforts should be made to benchmark as many available experimental properties as possible before performing ``true" computational predictions.

We considered the transferability of the \emph{U} values estimated in this work, with r$^2$SCAN, by examining systems for each TM with oxidation states and/or coordination environments not sampled while calculating the optimal \emph{U}. In general, we find that r$^2$SCAN or its Hubbard \emph{U} corrected version estimate similar lattice parameters and hence yield similar accuracies on structural properties. Analogously, the calculated on-site magnetic moments in SrFeO$_3$ and the band gaps in BiVO$_4$ are similar between r$^2$SCAN and r$^2$SCAN+\emph{U}. In case of electrochemical properties, we do find tangible variations in the calculated average voltages of r$^2$SCAN and r$^2$SCAN+\emph{U}, with r$^2$SCAN+\emph{U} exhibiting overall lower errors across the Co and Ni systems. Thus, we find the optimal \emph{U} values obtained in this work to be transferable across oxide frameworks not sampled \emph{a priori}. Nevertheless, more benchmarking studies to compare the performance of r$^2$SCAN+\emph{U} with r$^2$SCAN (and experiments) will help in quantifying the reliability and errors associated with using r$^2$SCAN+\emph{U}.

Given that r$^2$SCAN(+\textit{U}) is not systematically more or less accurate than SCAN(+\textit{U}), the computational performance and numerical stability of r$^2$SCAN(+\textit{U}) is critical in determining its utility in property predictions across materials. Thus, we have quantified the computational time of r$^2$SCAN(+\textit{U}) and SCAN+\textit{U} relative to SCAN for each TM-O$_2$ system considered in Figure~S1. Specifically, panels a, b, and c of Figure~{\ref{time}} plot the overall (electronic+ionic steps), per ionic step, and per electronic step computational time, respectively, taken by the SCAN+\textit{U} (blue bars), r$^2$SCAN (red), and r$^2$SCAN+\textit{U} (yellow) frameworks, relative to the computational time taken by the SCAN functional (dotted black lines), for each TM-based set of oxides. Details on calculating the computational times used by the functionals is described in the `Computational time' section of the SI. Note that our objective is not to provide a rigorous quantification of computational resources required for each XC framework, but to provide a qualitative understanding of the relative computational costs across the frameworks considered. 

For each electronic step, r$^2$SCAN(+\textit{U}) is typically faster than SCAN (Figure~\ref{time}), signifying better numerical stability than SCAN, with Mn, Ni, and Cu oxides being marginal exceptions. In contrast, on a per-ionic step basis, r$^2$SCAN and r$^2$SCAN+\textit{U} is slower than SCAN, by $\sim$1.05-1.78$\times$ and $\sim$1.1-1.31$\times$, respectively, highlighting that r$^2$SCAN(+\textit{U}) takes more electronic steps to converge per ionic step. Importantly, the overall computational time (ionic+electronic steps, Figure~{\ref{time}}) required for structural relaxation of TMOs using r$^2$SCAN and r$^2$SCAN+\textit{U} is lower than SCAN, by $\sim$12.1-61.2\% and $\sim$1.9-34.5\%, respectively, except in Fe oxides, indicating that r$^2$SCAN(+\textit{U}) takes lower number of ionic steps to converge, which possibly indicates a better description of atomic forces. The higher overall computation time in Fe oxides with r$^2$SCAN(+\textit{U}) than SCAN is primarily due to the difficulty in converging Fe$_3$O$_4$ with r$^2$SCAN(+\textit{U}). Comparing r$^2$SCAN and r$^2$SCAN+\textit{U}, we find that r$^2$SCAN+\textit{U} takes a higher overall computational time to converge, except in Fe and Ni oxides. Thus, we expect r$^2$SCAN(+\textit{U}) to provide good utility in property predictions in TM-containing systems given its better computational performance and reasonable accuracy compared to SCAN(+\textit{U}).

\section{Conclusion}
3\textit{d}-TMs and their compound phases find applications in several fields such as energy storage, solar cells, catalysts, thermochemical water splitting, etc., and it is imperative to predict their properties such as lattice constants, magnetic moments, reaction enthalpies, and band gaps accurately using DFT-based techniques for designing better materials. Recently, the r$^2$SCAN metaGGA XC functional was proposed to exhibit the accuracy of its predecessor, SCAN, and the computational performance of rSCAN in main-group compounds, but the accuracy of r$^2$SCAN was not rigorously tested on TM-based systems. Here, we assessed the numerical accuracy and computational performance of r$^2$SCAN in binary 3\emph{d}-TMOs, in calculating the lattice parameters, on-site magnetic moments, binary oxidation enthalpies, and band gaps against experimental data. Notably, we observed that r$^2$SCAN exhibited similar qualitative trends as that of SCAN, with marginally larger estimations of lattice parameters than SCAN, while the on-site magnetic moments and band gap calculations are marginally smaller than SCAN. While both r$^2$SCAN and SCAN underestimated the band gaps in wide gap TMOs, with SCAN offering slightly better accuracy, they failed to predict the correct ground state electronic configurations of narrow band gap TMOs (e.g., Mn$_2$O$_3$). 

On analysing the addition of Hubbard \textit{U}-correction to improve the accuracy of the r$^2$SCAN functional, we observed that a lower optimal \textit{U} value, based on experimental oxidation enthalpies, was required in a r$^2$SCAN+\emph{U} framework for Ti, Mn, Co and Ni oxides, when compared to a SCAN+\emph{U} framework. The optimal \emph{U} values were identical in both r$^2$SCAN+\emph{U} and SCAN+\emph{U} frameworks for V and Fe oxides, while we did not observe the need for a \emph{U} correction in Cr and Cu oxides with r$^2$SCAN, similar to SCAN. Moreover, introducing the \textit{U}-correction to SCAN and r$^2$SCAN increased the calculated lattice parameters, on-site magnetic moments and the band gaps of the TMOs. 

r$^2$SCAN+\textit{U} and SCAN+\textit{U} successfully opened a band gap for narrow gap TMOs (except VO$_2$ and Mn$_2$O$_3$ with r$^2$SCAN+\textit{U}). Upon testing the optimal \emph{U} values with r$^2$SCAN+\emph{U} on oxides with different oxidation states and/or coordination environments, we found that the \emph{U} values derived in this work are in general transferable to other TM-containing oxides as well. Furthermore, we observed that r$^2$SCAN(+\textit{U}) took less overall computational time (ionic+electronic steps) to converge when compared to SCAN, which indicated that r$^2$SCAN(+\textit{U}) was computationally more efficient than SCAN(+\textit{U}). Since r$^2$SCAN+\emph{U} offers a reasonably accurate prediction of material properties at a lower computational expense than SCAN+\emph{U}, we observe that r$^2$SCAN+\emph{U} can be used in high-throughput materials discovery, after adequate benchmarking tests are done in each new chemical space explored.\\

\subsection*{Acknowledgments}
G.S.G.\ acknowledges the Indian Institute of Science (IISc) Seed Grant, SG/MHRD/20/0020 and SR/MHRD/20/0013 and the Science and Engineering Research Board (SERB) of the Department of Science and Technology, Government of India, under sanction numbers SRG/2021/000201 and IPA/2021/000007 for financial support. R.D.\ thanks the Ministry of Human Resource Development, Government of India, for financial assistance. S.S.\ acknowledges financial support from SERB under IPA/2021/000007. All the authors acknowledge the computational resources provided by the Supercomputer Education and Research Centre, IISc, for enabling some of the density functional theory calculations showcased in this work.

\subsection*{Author Contributions} 
G.S.G. envisioned and designed the work. S.S. and R.D. performed the calculations. All authors contributed in data analysis and writing the paper.


\subsection*{Conflicts of Interest}
The authors declare no competing financial or non-financial interests.

\subsection*{Availability of data}
The data that support the findings of this study are openly available at \href{https://github.com/sai-mat-group/r2SCAN-U-benchmarking}{https://github.com/sai-mat-group/r2SCAN-U-benchmarking}.

\subsection*{Supplementary Materials}
Electronic Supporting Information is available online at , with details on the crystal structures used for calculations, oxidation energetics of Cr and Cu oxides, densities of states of all systems not showcased in the main text, and details on computational time calculations.


\bibliographystyle{unsrt}
\bibliography{sample}

\end{document}